# Emergence of superconducting dome in insulating ZrN$_x$ films via variation of nitrogen concentration

Fucong Chen[1,2], Xinbo Bai[1,2], Yuxin Wang[1], Tao Dong[3], Jinan Shi[2], Yanmin Zhang[1], Xiaomin Sun[4], Zhongxu Wei[1], Mingyang Qin[1], Jie Yuan[1,5], Qihong Chen[1,5], Xinbo Wang[1], Xu Wang[1], Beiyi Zhu[1], Rongjin Huang[4], Kun Jiang[1], Wu Zhou[2], Nanlin Wang[3], Jiangping Hu[1], Yangmu Li[1,2 *], Kui Jin[1,5, *], Zhongxian Zhao[1,5]

[1]Beijing National Laboratory for Condensed Matter Physics, Institute of Physics, Chinese Academy of Sciences, Beijing 100190, People's Republic of China
[2]School of Physical Sciences, University of Chinese Academy of Sciences, Beijing 100049, People's Republic of China
[3]International Center for Quantum Materials, School of Physics, Peking University, Beijing, China
[4]CAS Key Laboratory of Cryogenics, Technical Institute of Physics and Chemistry, Beijing 100190, China
[5]Songshan Lake Materials Laboratory, Dongguan, Guangdong 523808, People's Republic of China
correspondence to: yangmuli@iphy.ac.cn; kuijin@iphy.ac.cn

**Reproducing the electronic phase diagram of strongly correlated high-transition-temperature (high-$T_c$) superconductors in materials other than Cu-, Fe-, and Ni-based compounds has been a challenging task. We combine film growth, charge transport, magnetometry, Terahertz Spectroscopy, Raman scattering, and Scanning Transmission Electron Microscopy to investigate superconductivity and the normal state of ZrN$_x$, which reveals a phase diagram that bears extraordinary visual similarities to those of high-$T_c$ superconductors. By tunning the N chemical concentration, we observe the evolution of a superconducting dome in the close vicinity of a strongly insulating state and a normal state resistivity mimics its counterpart of the high-$T_c$ superconductors. Our detailed analyses demonstrate that the superconductivity of ZrN$_x$ can be characterized within the Bardeen-Cooper-Schrieffer paradigm and its normal state can be understood within the Fermi liquid framework. The visual similarities of the phase diagrams thus cannot be taken as an indicator of the same underlying physics.**

Previous studies have categorized most superconductors into two classes - the Bardeen-Cooper-Schrieffer (BCS) superconductors, for which the electron-phonon interactions drive the formation of Cooper pairs from a Fermi-liquid normal state [1], and the unconventional superconductors including the strongly correlated high-$T_c$ superconductors, where the electron-electron correlations dominate their pairing properties [2,3]. For typical high-$T_c$ superconductors, their parent compounds are Mott or charge transfer insulators. Upon chemical substitutions, a superconducting dome emerges close to the strongly insulating state, and a variety of intertwined charge and spin microscale patterns appear [4]. Their unconventional normal state is labeled by non-Fermi-liquid characteristics, such as the disappearance of part of the Fermi surface and occurrence of linear-in-temperature resistivity that holds up to unusually high temperatures [3,5]. The recent realization of strongly insulating states and superconducting dome in the twisted multi-layer Graphene [6–8] and van der Waals materials [9] suggests that certain features of unconventional superconductivity may not be limited to 3$d$ transition metal compounds such as Cu- [3,5], Fe-, [10,11] and Ni-based superconductors [12]. Searching for other material systems that present a similar phase diagram can significantly advance our understanding of the superconductivity.

Superconductivity in transition metal nitrides was established in the 1950s, and the nature of their superconductivity has been investigated ever since [13–19]. Early studies reported tunable superconductivity using nitrogen concentrations [18,19]. For most of these materials, their superconducting properties, including superconducting gaps and upper critical fields, can be explained by the BCS physics [20]. However, a few early works also reported experimental observations that require theory beyond the BCS picture, e.g., the density of states at the Fermi surface [21] and electron-phonon coupling strength [22] are too low to account for their relatively high $T_c$. Meanwhile, because of their mechanical hardness and durability, superconducting transition metal nitrides have attracted a lot of attention lately for their unique advantages in technological applications [23–26].

In this letter, we study the archetypical transition metal nitrides, ZrN$_x$, with a range of experimental techniques. Combining film growth, charge transport, magnetometry, Terahertz Spectroscopy, Raman scattering, and Scanning Transmission Electron Microscopy, we construct a phase diagram of ZrN$_x$ in detail. We found the emergence of superconductivity close to a strongly insulating (SI) state. The superconducting dome, which is controlled by the N concentration, evolves from the SI state at large $x$ to a weakly insulating (WI) state at small $x$. Above the superconducting dome and between the two insulating states, a metallic normal (M) state exists with linear-in-temperature resistivity up to almost room temperature. The overall phase diagram of ZrN$_x$ thus resembles those for the high-$T_c$ superconductors.

ZrN has a NaCl structure with lattice constants of ~ 4.60 Å, which belongs to the $Fm\bar{3}m$ space group and the $O_h$ point group. ZrN$_x$ studied previously often have multiple structure domains, e.g., in refs [14,27], which renders a comprehensive study of ZrN$_x$'s intrinsic properties difficult. Here, we grew (00$l$) oriented ZrN$_x$ films on (00$l$) MgO substrate with thicknesses of ~ 150 nm using Pulsed Laser Deposition. By employing fine controls of nitrogen partial pressure and the relative position between the substrate and target, ZrN$_x$ films with $x$ from 0.54 to 1.40 were obtained. The N concentration and crystal lattice parameters were measured with Energy Dispersive X-Ray Spectroscopy (EDX, Fig. S1) and X-ray diffraction (XRD, Fig. S2). All ZrN$_x$ thin films studied in this work have the same ZrN $Fm\bar{3}m$ structure (Table S1).

Figure 1 illustrates the temperature-dependent resistivity of ZrN$_x$ films. For $x \sim 1.00$, the onset superconducting transition temperature, $T_c^{\text{onset}}$, reaches a maximum value of 10 K, consistent with that of the bulk ZrN$_x$ [13]. For the N-rich region ($x > 1.00$), $T_c^{\text{onset}}$ gradually decreases towards zero and a strongly insulating phase develops. The resistivity rises by two orders of magnitude from 300 K to 2 K [Fig. 1(a)], and a resistivity upturn appears [Fig. 1(b)].

Notably, the onset superconducting transition is clearly seen for insulating ZrN$_x$ with $1.30 \leq x \leq 1.35$. For the N-deficient region ($x < 1.00$), a decrease of $T_c^{\text{onset}}$ coincides with the presence of a weakly insulating phase, for which the resistivity upturn is much less obvious [Fig. 1(b)]. Representative temperature-dependent magnetic susceptibility of ZrN$_x$ is delineated in Fig. 1(c). The superconducting volume fraction is found to be close to 1, suggesting bulk rather than surface superconductivity. We performed measurement of the superconducting energy gap, $2\Delta$, using Terahertz spectroscopy (see Terahertz raw data in Fig. S3). As plotted in Fig. 1(d), the temperature dependence of $2\Delta$ can be well explained by the BCS gap function, $2\Delta/k_B T_c \sim (1 - \frac{T}{T_c})^{1/2}$ [1], implying that ZrN$_x$ is a BCS superconductor. In Fig. 1(e), we construct a phase diagram for ZrN$_x$. Strikingly, the superconducting dome emerges in the close vicinity of the SI phase and a metallic (M) state exists above $T_c$ near the optimal N concentration.

Charge transport properties of the normal state are presented in Fig. 2. In Fig. 2(a) and (b), we plot resistivity of ZrN$_x$ films as functions of $T$ and $T^2$, respectively. Interestingly, the longitudinal resistivity, $\rho_{xx}$, can be described by linear-in-temperature and quadratic-in-temperature regimes above the superconducting dome. This resistivity behavior, $\rho_{xx} = \rho_0 + A_1 T + A_2 T^2$, where $A_1$ and $A_2$ are resistivity prefactors and $\rho_0$ is the residual resistivity, imitates resistivity of the high-$T_c$ superconductor normal state [3,28–33]. For quantitative analysis, we take the first derivative of $\rho_{xx}$. Fig. 2(c) presents representative $d\rho_{xx}/dT$ data of a ZrN$_x$ film, for which $\rho_{xx} \propto A_1 T$ holds above temperature $T_1$ and $\rho_{xx} \propto A_2 T^2$ below $T_2$. We superimpose the temperature dependence of resistivity on top of the phase diagram as a pseudo-color plot in Fig. 2(d). Alternatively, the electrical resistivity of ZrN$_x$ can be analyzed based on the Matthiessen rule with electron-phonon scattering, where the total resistivity was written as $\rho_{xx} = \rho_0 + \rho_{\text{ph}}$. $\rho_0$ again is the residual resistivity and $\rho_{\text{ph}} \sim T^n$ is the electron-phonon scattering term [20]. For temperature far below the Debye temperature $\rho_{\text{ph}} \sim T^5$, and for temperatures comparable to the Debye temperature $\rho_{\text{ph}}(T) \sim T$. In Fig. S4, we carried out the additional fit of $\rho_{xx}$, using $\rho_{xx} = \rho_0 + \rho_{\text{ph}}$ and found that electron-phonon scattering describes our resistivity data equally well. Thus, the similarities of the normal state transport between ZrN$_x$ and high-$T_c$ superconductors cannot be taken as evidence for non-Fermi-liquid physics in ZrN$_x$.

Hall coefficients of ZrN$_x$ with magnetic fields perpendicular to the film from 9 T to -9 T are depicted in Fig. 3. $\rho_{xy}$ is proportional to $H$ and for each N concentration it collapses onto each other for a wide temperature range. Hall coefficient $R_H$ is insensitive to the temperature, which is indicative of a relatively simple temperature-independent Fermi surface. In Fig. S5, we calculated the band structure and Fermi surface for ZrN using density-functional theory (DFT), employing the projector augmented wave method and the generalized gradient approximation [34,35]. We observe the existence of electron Fermi pockets that are consistent with the negative sign of the Hall resistivity.

Raman spectroscopy has been carried out to characterize the ZrN$_x$ phonon modes at room temperature. Theoretically, due to the $O_h$ point symmetry of the ZrN structure, the first-order Raman scattering should be prohibited [36]. However, the excess/absent N atoms and disorders can destroy the symmetry protection, leading to the rise of acoustic and optical phonon Raman intensities [36,37]. Figure. 4(a) illustrates the Raman spectra of ZrN$_x$ with various $x$ (detailed Raman data in Fig. S6). We convert the Raman shift of acoustic phonon modes to the electron-phonon coupling constant $\lambda \propto 1/\omega^2_{acoustic}$ in Fig. 4(b) [37], which reveals a significant weakening of $\lambda$ in the N-deficient region. Because both Raman intensities and residual resistivity $\rho_0$ can reflect the amount of disorder in ZrN$_x$, we compare them in Fig. 4(c), which display a large enhancement in the SI phase. To further investigate the disorder in the SI phase, we performed Scanning Transmission Electron Microscopy (STEM) measurements on ZrN$_x$ with $x = 1.00, 1.25, 1.35$. STEM data at the atomic scale for $x = 1.25$ reveal additional N atoms located between the Zr atoms (also see Fig. S7 for STEM data with other $x$). We summarize the evolution of the superconducting dome in Fig. 4(d). At $x = 1.00$, ZrN$_x$ has an optimal $T_c$ of 10 K. With a lowering $x$, the electron-phonon coupling obtained from Raman scattering decreases. Applying McMillan's formula, we estimate the upper bound of the superconducting transition temperature, $T_c^* = (\theta_D/1.45)\exp[-\frac{1.04(1+\lambda)}{\lambda-\mu^*(1+0.62\lambda)}]$, where $\theta_D$ is the Debye temperature (~ 475 K from fit, consistent with that in [20]), $\mu^*$ is the Coulomb repulsion strength (~ 0.1 from fit) [38,39]. In the N-deficient region, the calculated $T_c^*$ tracks the experimentally obtained $T_c^{onset}$ very well. However, for the N-rich region, $T_c^*$ is almost a constant, and thus cannot explain the lowering $T_c^{onset}$. Alternatively, assuming the presence of excess N atoms [Fig. 4(e)] can induce localization of otherwise mobile electrons due to bonding between negatively charged N and positively charged Zr atoms, we estimate the number of mobile electrons per Zr atoms, valence $n_e$/Zr, as a function of $x$ in Fig. 4(d) (dashed line). The value of $n_e$/Zr $\sim 4 - 4/3x$ for $x < 1.33$ reflects how many electrons can contribute to the normal state and superconductivity, and it is found to be consistent with that extracted from Hall measurement for $x > 1.00$. We note that Zr$_3$N$_4$ was previously calculated to be a band insulator [40]. Our simple estimation here only based on the electron valency of Zr and N, with no requirement of the Zr$_3$N$_4$ structure (all our ZrN$_x$ films are characterized to be the ZrN $Fm\bar{3}m$ structure).

In the N-deficient region, because the N to Zr ratio is less than one, our estimation based on excess N atoms no longer works. We performed DFT calculation with virtual crystal approximation. The DFT $n_e$/Zr is calculated by integrating density of states for Zr bands up to the energy where the integral of all bands equals to the number of carriers in the material. We compare the DFT calculated $n_e$/Zr to that extracted from Hall measurement and obtain a good correspondence for $x < 1.00$.

The insulator-to-superconductor transition near the SI phase and the evolution of superconducting dome as a function of N concentration in ZrN$_x$ mimic what happens in high-$T_c$ superconductors visually. Such a resemblance inevitably raises questions about the underlying physics regarding the two systems. For high-$T_c$ superconductors, large onsite Coulomb repulsion results in Mott or charge transfer insulators that coexists with long-range antiferromagnetism [3-5]. In the case of ZrN$_x$, our magnetic susceptibility measurements observe no magnetic orderings inside and outside of the SI phase. STEM results (Fig. S7) demonstrate the appearance of interstitial N atoms when ZrN$_x$ enters the SI phase, while no ordering of the interstitial N atoms has been observed. The drastic increase of resistivity and decrease in the number of carriers can thus be attributed to charge localization associated with additional N atoms. In principle, the reduction of mobile electrons would limit the number of Cooper pairs and lead to a lower $T_c$. In fact, merely based on the valency of Zr and N atoms, we estimate the upper superconducting dome boundary at $x = 1.33$, where $n_e/\text{Zr} \sim 4 - 4/3x = 0$. This estimation is consistent with experimental observed values in the N-rich region [Fig. 4(d)]. For the N-deficient region, $T_c^{\text{onset}}$ is found to follow $T_c^*$, suggesting that the electron-phonon coupling dominates the evolution of the superconducting dome. These results point to a BCS-like superconductivity of $ZrN_x$.

The nature of the metallic normal state for $ZrN_x$, which exists between the SI and WI phases can be understood based on its charge transport properties. Although no consensual conclusion has been made on the underlying physics of the normal state for high-$T_c$ superconductors [3, 41], relevant main signatures include linear-in-temperature resistivity that goes above the Mott-Ioffe-Regel limit and a temperature-dependent Hall coefficient [3,28–33]. The scaling relation of resistivity prefactors was also linked to possible existence of quantum criticality and the emergence of superconductivity [29,33,42]. For $ZrN_x$, we only observe $\rho_{xx} \sim T$ resistivity near 250 K, almost half of the Debye temperature, while the resistivity magnitude is well below the Mott-Ioffe-Regel limit (Table S2). The $\rho_{xx} \sim T$ resistivity is thus more likely to associate with electron phonon scattering. The N concentration dependence of $A_1$, and $A_2$ also show no obvious scaling relation respective to $x$ [Fig. 2(d) and Fig. S4], and they are closely related to $\rho_0$, that is suggestive of overall enhancement of resistivity instead of unconventional charge transport features. The nearly temperature-independent Hall coefficient, in addition, provides clear evidence for the persistence of a full Fermi surface, in contrast to that for high-$T_c$ superconductors. Recently, the linear-in-temperature resistivity in high-$T_c$ superconductors was found to be associated with the Planckian scattering rate, $\hbar/\tau \sim \alpha k_B T$, where $\alpha$ is close to 1 [43,44]. We estimate that the temperature range at which $\rho_{xx} \propto A_1 T$ appears for $ZrN_x$ is also much lower than the Planckian limit (Table S2). Thus, the resistivity of $ZrN_x$ likely originates from an electron-phonon scattering of a Fermi-liquid normal state that relevant to excess and vacant N atoms. No evidence of Mott physics or strong electron-electron correlation has been found for $ZrN_x$.

$ZrN_x$ film with tunable nitrogen concentration presents a remarkable material system, for which the electronic phase diagram looks astonishingly similar to that of the high-$T_c$ superconductors [3,5,10,11], even though its key characteristics of superconductivity and normal state seemingly require no exotic theory to explain. In light of recent studies on the phase diagram of high-$T_c$ superconductors [12,33,43,44] and attention given to the strongly correlated physics in multi-layer Graphene and related systems [6-8], our findings call for caution in assessing the underlying physics based solely on the similarity between phase diagrams. Furthermore, through the detailed study of $ZrN_x$, we demonstrate a well-controlled chemical tunning of the superconducting and normal states for the technologically useful transition metal nitrides.

Acknowledgement: We thank Lihong Yang, Hua Zhang, Xiaoli Dong, Huaixin Yang, and Zhanyi Zhao for assistant in experimental measurements and Jian Kang for useful discussions.

This work was supported by the National Key Basic Research Program of China (2021YFA0718700, 2017YFA0302900, 2017YFA0303003, 2018YFB0704102, and 2018YFA0305800), the National Natural Science Foundation of China (11888101, 11927808, 11834016, 11961141008 and 12174428), the Strategic Priority Research Program (B) of Chinese Academy of Sciences (XDB25000000, XDB33000000), CAS Interdisciplinary Innovation Team, Beijing Natural Science Foundation (Z190008), Key-Area Research and Development Program of Guangdong Province (2020B0101340002) and the Center for Materials Genome.

References:

[1] J. Bardeen, L. N. Cooper, J. R. Schrieffer, Phys Rev, **108**(5) (1957).
[2] D. J. Scalapino, Rev. Mod. Phys. **84**, 1383 (2012).
[3] B. Keimer et al., Nature **518**, 179 (2015).
[4] E. Fradkin, S. A. Kivelson, and J. M. Tranquada, Rev. Mod. Phys. **87**, 457 (2015).
[5] N. P. Armitage, P. Fournier, and R. L. Greene, Rev. Mod. Phys. **82**, 2421 (2010).
[6] Y. Cao et al., Nature **556**, 43 (2018).
[7] Y. Cao et al., Nature **556**, 80 (2018).
[8] G. Chen et al., Nature **572**, 215 (2019).
[9] J. T. Ye et al., Science **338** 1193 (2012).
[10] Y. Mizuguchi and Y. Takano, J. Phys. Soc. Jpn. **79**, 102001 (2010).
[11] P. Dai, Rev. Mod. Phys. **87**, 855 (2015).
[12] D. Li et al., Nature **572**, 624 (2019).
[13] B. T. Matthias and J. K. Hulm, Phys. Rev. **87**, 799 (1952).
[14] R. S. Ningthoujam and N. S. Gajbhiye, Prog. Mater. Sci. **70**, 50 (2015).
[15] Y. Guo et al., J. Am. Chem. Soc. **141**, 10183 (2019).
[16] S. Yamanaka, K. ichi Hotehama, and H. Kawaji, Nature **392**, 580 (1998).
[17] Y. Nakagawa et al., Science **372**, 190 (2021).
[18] W. Lengauer, Surf. Interface Anal. **15**, 377 (1990).
[19] R. E. Treece, J. S. Horwitz, and D. B. Chrisey, Mater. Res. Soc. Symp. Proc. **343**, 747 (1994).
[20] A. Cassinese, M. Iavarone, R. Vaglio, M. Grimsditch, and S. Uran, Phys. Rev. B **62**, 13915 (2000).
[21] H. Tou, Y. Maniwa, T. Koiwasaki, and S. Yamanaka, Phys. Rev. Lett. **86**, 5775 (2001).
[22] Y. Kasahara, K. Kuroki, S. Yamanaka, and Y. Taguchi, Physica C **514**, 354 (2015).
[23] F. Rivadulla, M et al., Nat. Mater. **8**, 947 (2009).
[24] B. Anasori, M. R. Lukatskaya, and Y. Gogotsi, Nat. Rev. Mater. **2**, 16098 (2017).
[25] A. Zerr, G. Miehe, and R. Riedel, Nat. Mater. **2**, 185 (2003).
[26] M. Chhowalla and H. E. Unalan, Nat. Mater. **4**, 317 (2005).
[27] Y. Yuan et al., Nat. Mater. **19**, 282 (2020).
[28] C. Proust and L. Taillefer, Annu. Rev. Condens. Matter Phys. **10**, 409 (2019).
[29] R. A. Cooper et al., Science **323**, 603 (2009).

[30] K. Jin, N. P. Butch, K. Kirshenbaum, J. Paglione, and R. L. Greene, Nature **476**, 73 (2011).
[31] Y. Li, W. Tabis, G. Yu, N. Barišic, and M. Greven, Phys. Rev. Lett. **117**, 197001 (2016).
[32] N. Barisic *et al.*, Proc. Natl. Acad. Sci. USA **110**, 12235 (2013).
[33] J. Yuan *et al.*, Nature in press, arXiv 2103.0835 (2021).
[34] G. Kresse and J. Hafner, Phys. Rev. B **47**, 558 (1993).
[35] J. P. Perdew, K. Burke, and M. Ernzerhof, Phys. Rev. Lett. **78**, 1396 (1997).
[36] W. Spengler and R. Kaiser, Solid State Commun. **18**, 881 (1976).
[37] X. J. Chen *et al.*, Phys. Rev. B **70**, 014501 (2004).
[38] W. L. McMillan, Phys. Rev. **167**, 331 (1968).
[39] J. Geerk, G. Linker, and R. Smithey, Phys. Rev. Lett. **57**, 3284 (1986).
[40] W. Y. Ching, Y. Xu, and L. Ouyang, Phys. Rev. B **66**, 235106 (2002).
[41] M. R. Norman, Science **332**, 196 (2011).
[42] S. Badoux *et al.*, Nature **531**, 210 (2016).
[43] G. Grissonnanche *et al.*, Nature **595**, 667 (2021).
[44] J. Ayres *et al.*, Nature **595**, 661 (2021).

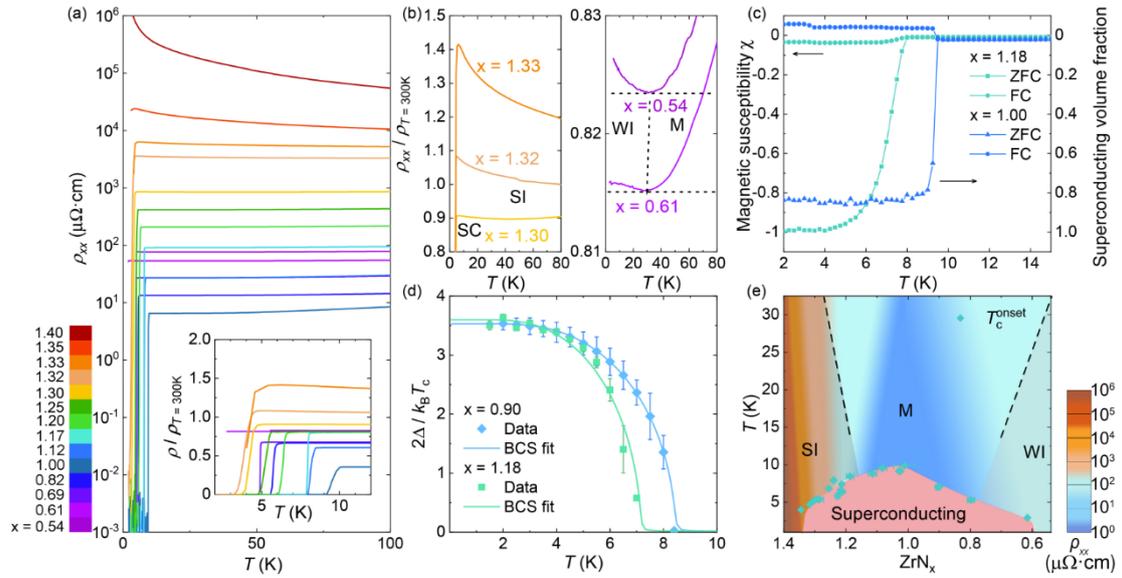

**Fig. 1. Superconductivity of ZrN$_x$ films. (a) Temperature-dependent resistivity of ZrN$_x$ with various $x$. The inset shows the superconducting transition. (b) Resistivity upturns in the N-rich (left panel) and N-deficient (right panel) regions. SC, SI, WI, and M corresponds to the superconducting, strongly insulating, weak insulating, and metallic phases, respectively. (c) Representative magnetic susceptibility and superconducting volume fraction. (d) Temperature dependence of superconducting gap, 2Δ, measured by terahertz spectroscopy. Error bars estimated based on the fits using the Mattis-Bardeen formula. (e) Phase diagram of ZrN$_x$. The color shades outside of the superconducting dome indicate the magnitude of electrical resistivity.**

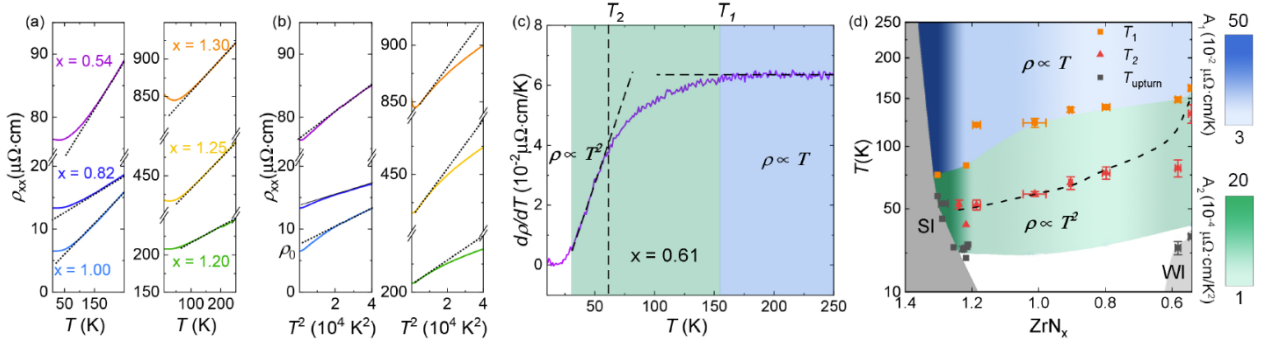

Fig. 2. Electrical resistivity in the metallic state. (a) Linear-in-temperature resistivity. Dashed line are the fits to $\rho_{xx} \propto A_1 T$ at high temperatures. (b) Quadratic-in-temperature resistivity. Dashed line are the fits to $\rho_{xx} \propto A_2 T^2$ at low temperatures. We extrapolate the zero-temperature intercept as $\rho_0$. (c) First order derivative of resistivity respect to temperature. $T_1$ and $T_2$ are characteristic temperatures for $\rho_{xx} \propto T$ and $\rho_{xx} \propto T^2$, respectively. (d) Normal state phase diagram of ZrN$_x$ based on electrical resistivity. Color shades indicate the resistivity prefactors $A_1$ and $A_2$ obtained from data in (a) and (b). Error bars of characteristic temperatures estimated based on the fits of resistivity.

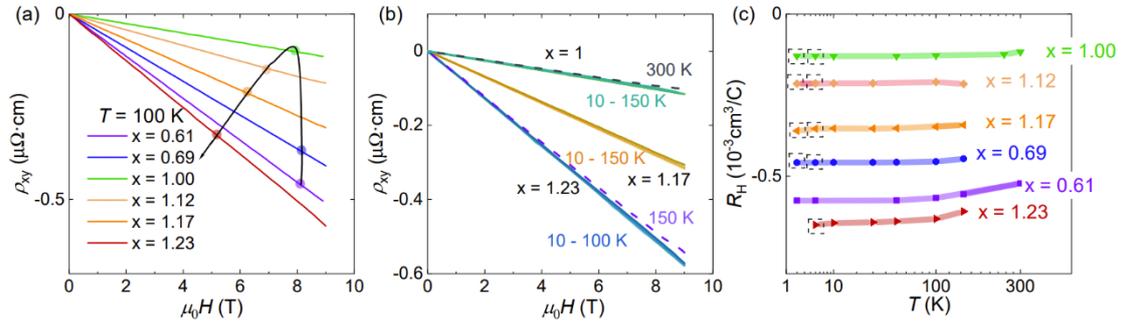

Fig. 3. Hall effect of ZrN$_x$ films. (a) Representative Hall resistivity, $\rho_{xy}(H)$, for different $x$ at 100 K. (b) $\rho_{xy}(H)$ for $x$ =1, 1.17 and 1.23 from 10 K to the highest measured temperatures. (c) The Hall coefficient, $R_H$, as a function of N concentration and temperature. Dotted boxes indicate Hall coefficients obtained at low temperatures when superconductivity is suppressed. Error bars of $R_H$ are smaller than the symbol size.

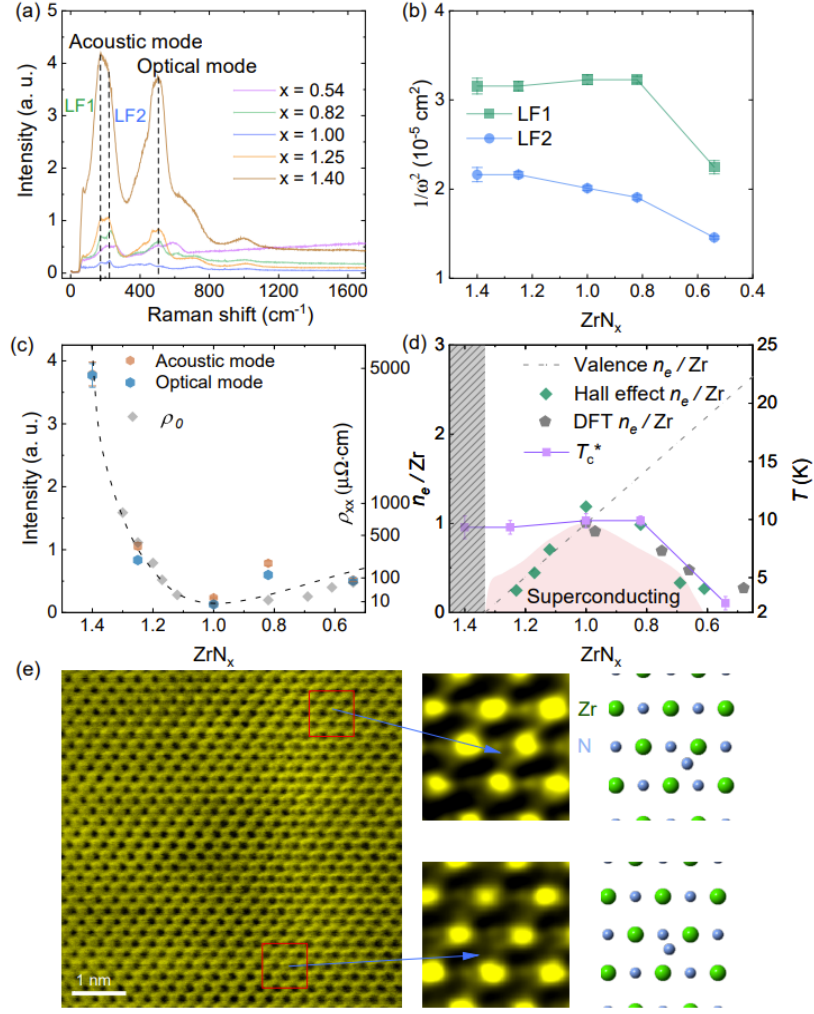

**Fig. 4. Raman scattering and Scanning Transmission Electron Microscopy results.** (a) Raman spectra of ZrN$_x$ with various $x$. LF1 and LF2 correspond to the two low-frequency acoustic phonon modes [35]. (b) Inverse square of the acoustic mode Raman shift, $1/\omega_{ac}^2$, for LF1 and LF2, which is proportional to the electron-phonon coupling. (c) Comparison between Raman scattering intensities and the residual resistivity, $\rho_0$. (d) Analysis for superconducting dome. As described in the text, valence $n_e/\text{Zr}$ and DFT $n_e/\text{Zr}$ are number of estimated mobile electrons per Zr for N-rich and N-deficient regions, respectively. Hall effect $n_e/\text{Zr}$ are the experimental data. Upper bound of the superconducting transition temperature, $T_c^*$, is calculated from $1/\omega_{ac}^2$ based on the McMillan's formula [35]. Dashed grey shade for $x > 1.33$ signifies an upper boundary for superconductivity. Error bars of Hall effect $n_e/\text{Zr}$ are smaller than the symbol size. (e) STEM image of ZrN$_x$ for $x = 1.25$. Additional N atoms (blue in the illustration) are revealed to locate between the Zr atoms (Green).

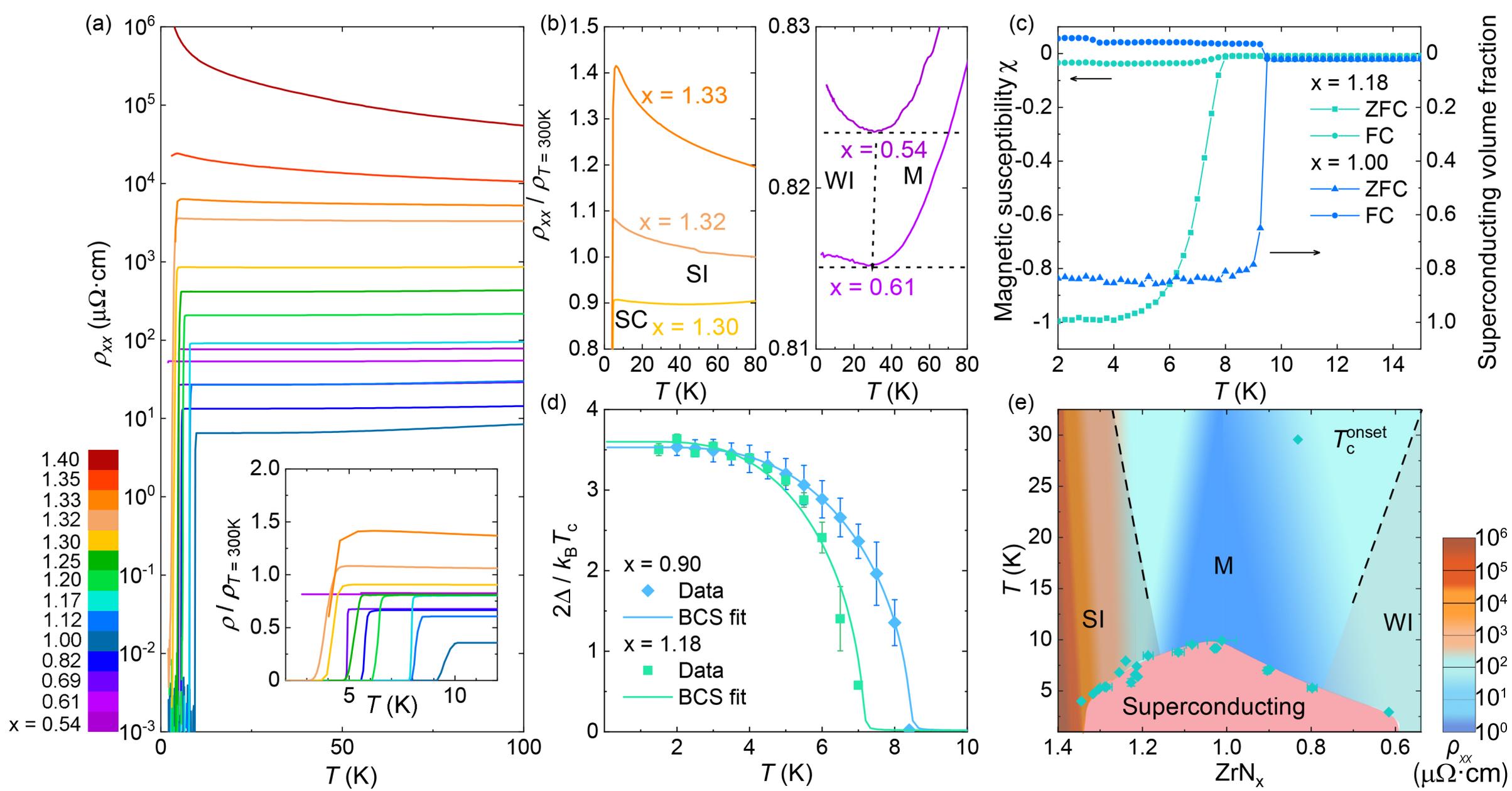

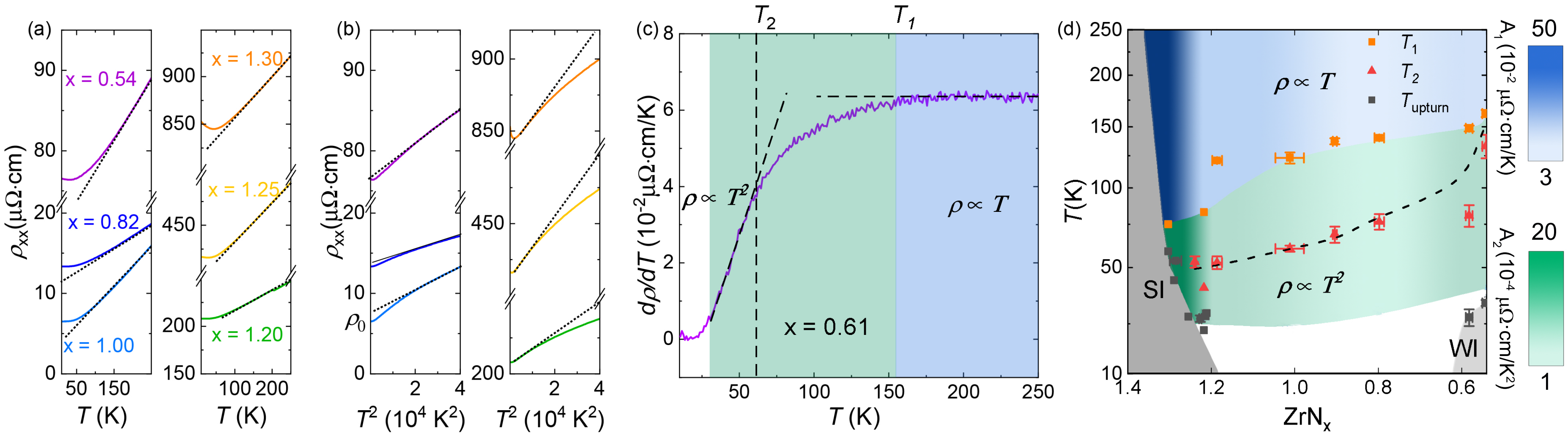

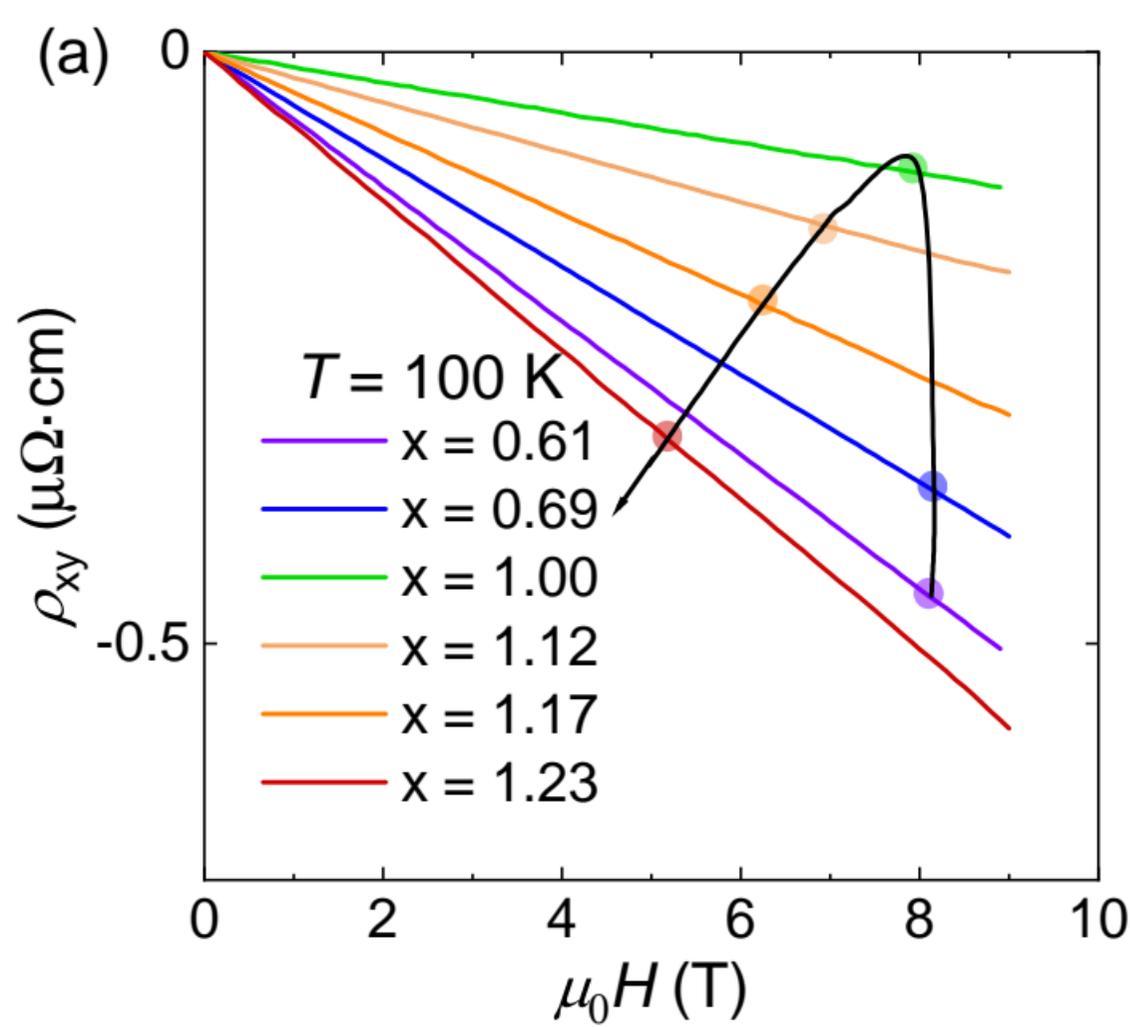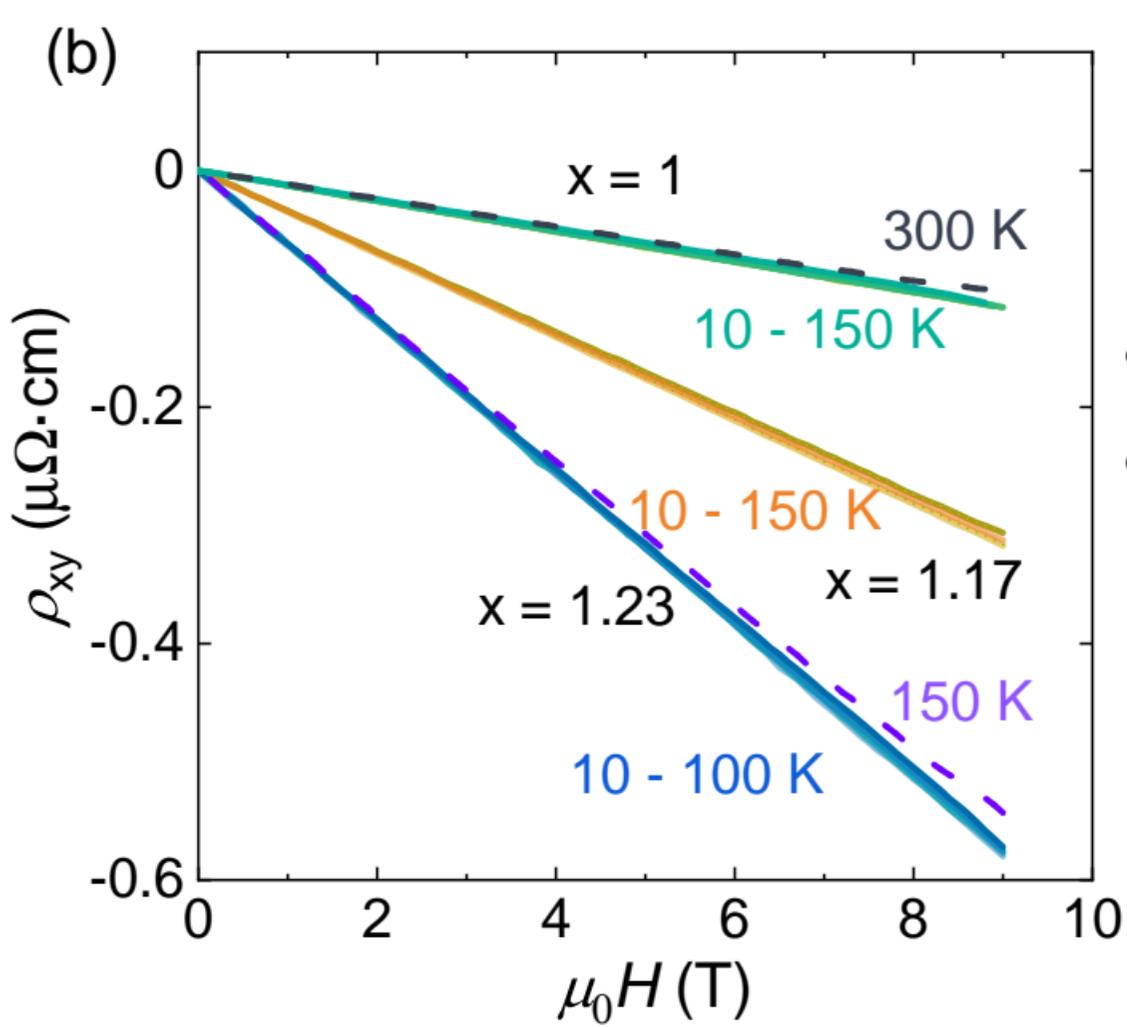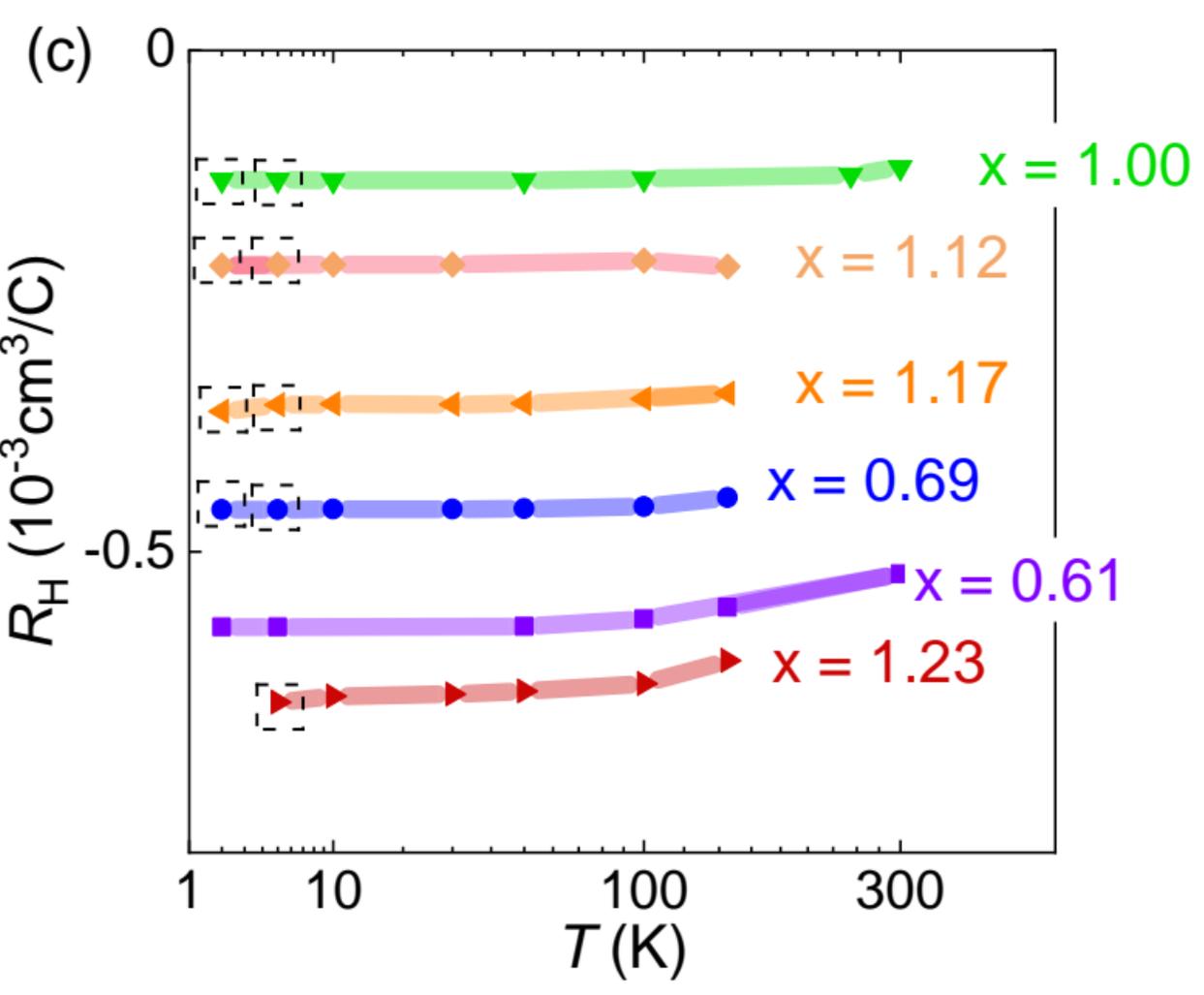

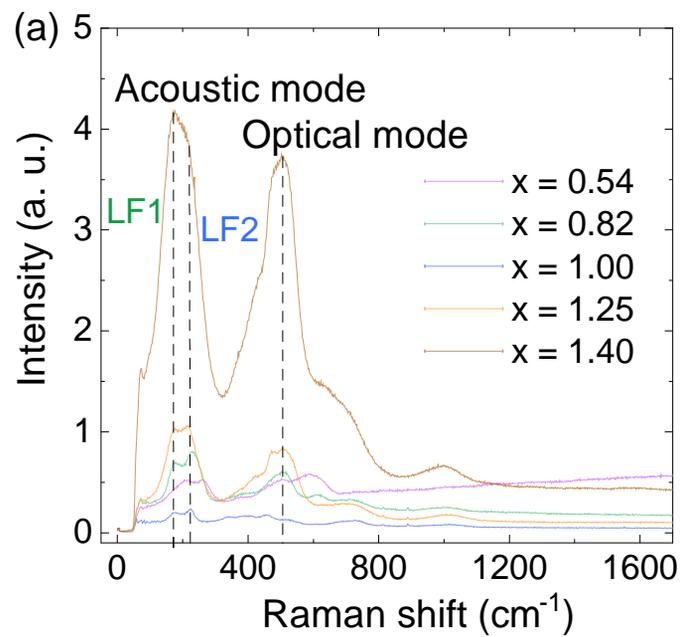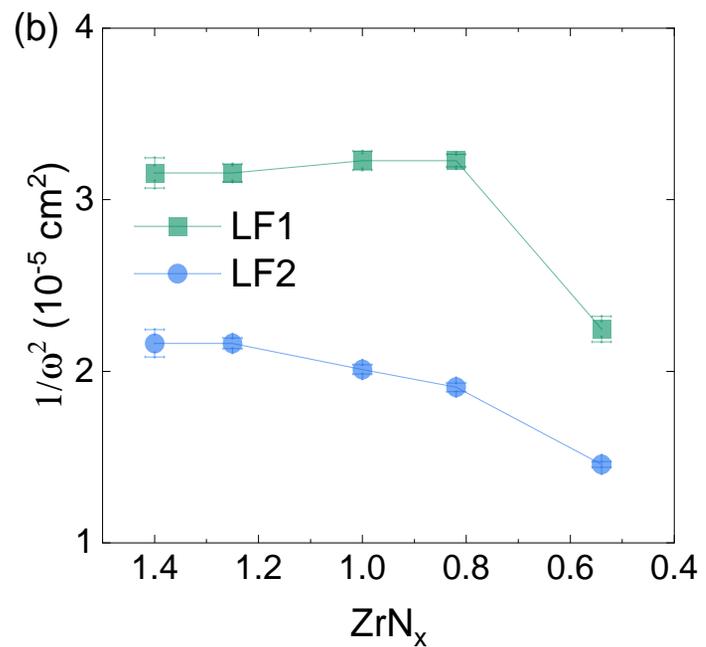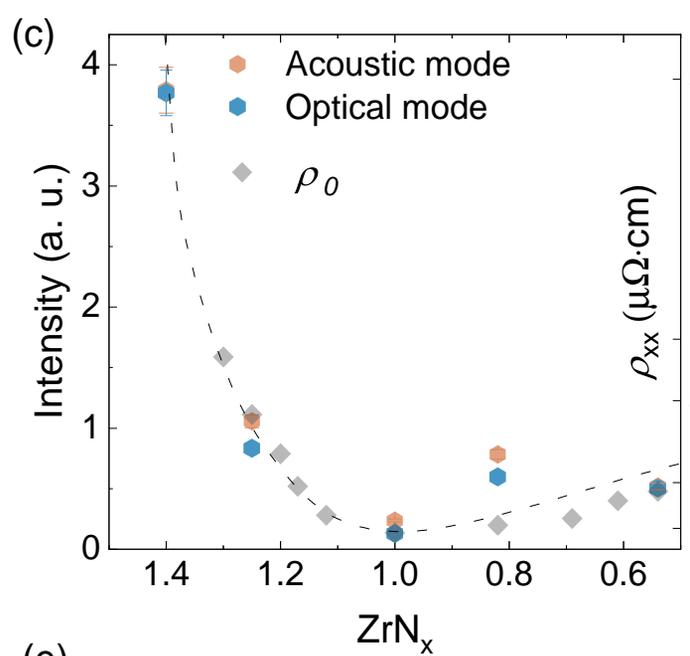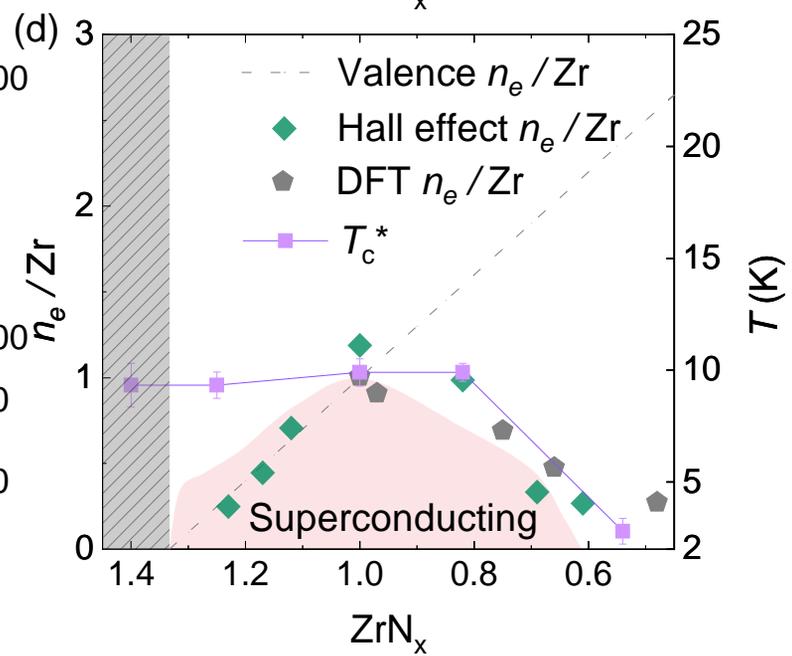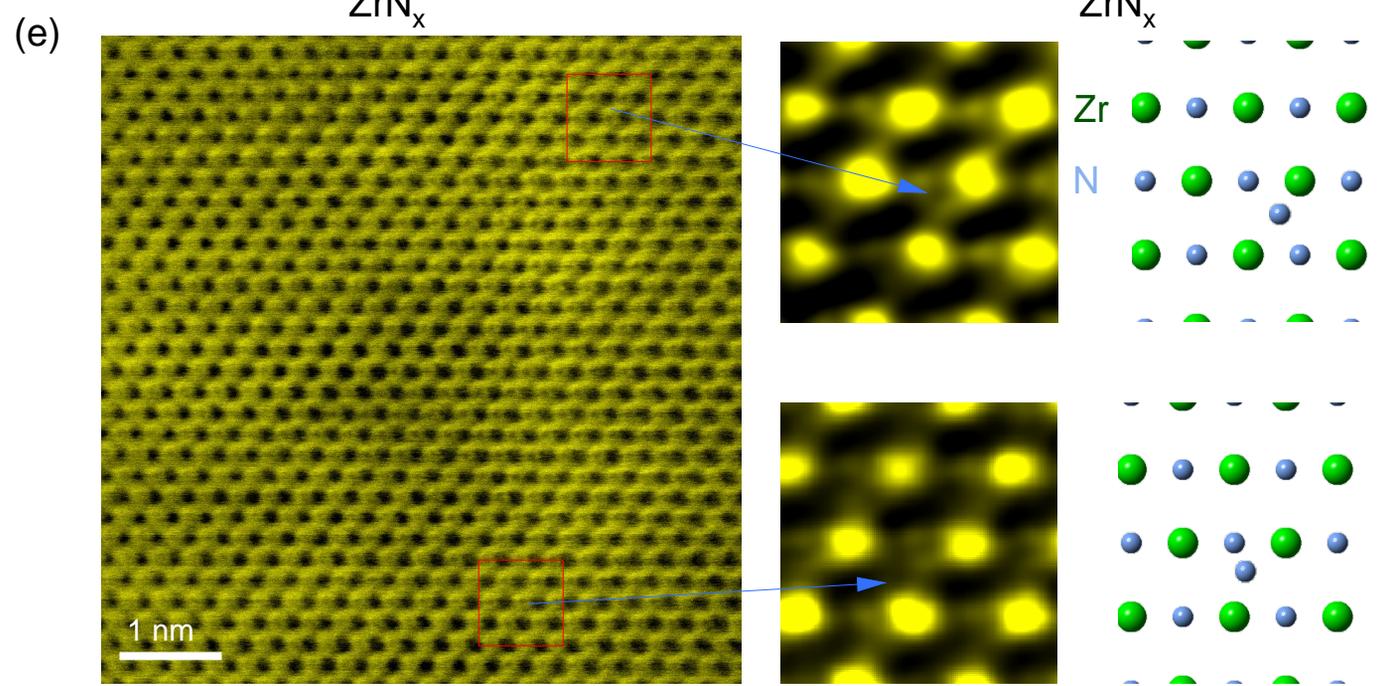